\documentclass[preprint,amsmath,amssymb,aps,]{revtex4-2}

\usepackage{graphicx}
\usepackage{dcolumn}
\usepackage{bm}

\begin{document}

\title{Optical microscope with nanometer longitudinal resolution based on a Linnik interferometer}

\author{Sergei V. Anishchik}
\email[E-mail: ]{anishchik01@gmail.com}
\affiliation{Department of Chemistry, Michigan State University, East Lansing, Michigan 48824, USA}

\author{Marcos Dantus}
\email[E-mail: ]{dantus@chemistry.msu.edu}
\affiliation{Department of Chemistry, Michigan State University, East Lansing, Michigan 48824, USA}
\affiliation{Department of Physics and Astronomy, Michigan State University, East Lansing, Michigan 48824, USA}

\begin{abstract}
A microscope based on the Linnik interferometer was designed, built, and tested. Two methods were used for interference pattern measurement: phase-shifting and polarized single-shot methods. The former uses a low coherence light emitting diode as a light source, providing 10 nm resolution in the Z direction and diffraction-limited resolution in the X and Y directions. The second method is insensitive to vibrations and enables observation of moving objects. The simplicity and low cost of this instrument make it valuable for a variety of applications.
\end{abstract}

\maketitle

\section{Introduction}

Light interference is widely used in science and technology \cite{Malacara2005,Malacara2007,Wyant2018,Fringe2013,deGroot2022}. Unlike most optical instruments, where useful information is contained in the intensity of the light, interferometry extracts the necessary information from the phase of the light wave.  This made it possible to achieve picometer resolutions in the optical shop testing and optical microscopy \cite{Saif2017,Minaev2018} and absolutely fantastic resolutions that made it possible to detect gravitational waves \cite{Abbott2016,Abbott2017}.

Our goal was to design and build a simple 3D super-resolution microscope that could be used to observe a wide variety of objects. We used  the scheme of the Linnik interferometer \cite{Linnik1933}.  In profilometry, the Mirau interferometer \cite{Mirau1952} is very popular because it allows easy automation of the measurement process. However, the Linnik interferometer, due to its symmetry,  is very easily modified, uses conventional objective lenses, and provides higher spatial resolution.

Various methods are used to decode interferograms, such as the Fourier transform \cite{Malacara2005,Malacara2007,Takeda1983} and the Hilbert transform \cite{Wang2013}. But in our case, it turned out to be more reliable to use the phase-shifting interferometry method \cite{Malacara2007,Wyant1984,Farrell1992,Wyant2011}, for which a lot of work has been done on error elimination methods \cite{Schwider1983,Hariharan1987,Schmit1995,Schmit1996}. The phase-shifting method is simple to implement and produces high-quality images. It allows the use of low-coherence light, which eliminates speckles and interference not directly related to the sample. However, it is not suitable for observing moving objects due to its long measurement time. Additionally, it is crucial to eliminate any vibration and drift.

Single-shot methods enable quick acquisition of an interference pattern, exhibit low sensitivity to vibrations, and facilitate observation of object motion. One of the methods for this involves using multiwavelength illumination \cite{Nakata2013}. However, better results were obtained with the method based on the manipulation of light polarization \cite{Wyant2018,Millerd2001,Novak2005,Millerd2005,Brock2005,Millerd2006,Millerd2007,Hong2024}. Therefore, we used the latter in our work.

\section{Methods for interference pattern measurement and extraction of sample structure information}

\subsection{Technique for analyzing the surface of a sample using an interference pattern with a polarization-insensitive camera and a low-coherence light source}

\begin{figure}[b]%[ht!]
\centering\includegraphics[width=0.95\textwidth]{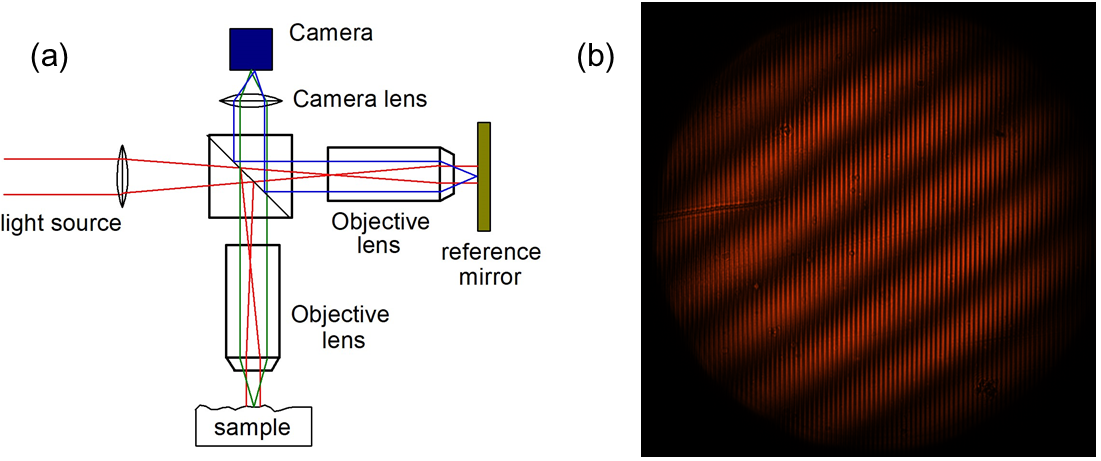} \caption{(a) Scheme of Linnik interferometer. (b) Interference fringe pattern of a reflective grating with 600 lines/mm. \label{Fig1}} 
\end{figure}

\subsubsection{Construction of the microscope}

Fig.~\ref{Fig1}(a) displays the main diagram of our microscope, which operates based on the Linnik interferometer. The light from the source passes through a beam splitter and reaches the sample and reference mirror. The reflected light enters through the same beam splitter onto the light-sensitive matrix of the camera, where interference of the two light rays occurs. Our microscope's operating principle is similar to that of the Twyman-Green interferometer \cite{Malacara2007}. The Linnik interferometer uses identical objective lenses in both arms of the interferometer, providing several significant advantages. These advantages will be discussed later.

For most experiments a LED  (M595F2, Thorlabs) with a center wavelength $\lambda_0$ near 595~nm and a bandwidth with full-width at half maximum $\Delta \lambda$ of about 70~nm was used as the light source. Thus, the coherence length $\lambda^2_0/\Delta\lambda $ was about 5~$\mu$m. This eliminates the presence of speckles, but places strict requirements on the symmetry of the interferometer arms. All results in our paper were obtained using air objectives with magnification 60x and NA = 0.9. However, we have used other types of objective lenses, including immersion lenses, without any problems. The reference mirror was equipped with a piezo stage with a positioning accuracy of 1~nm, which is necessary for using of phase-shifting method. The spatial resolution of the microscope is highly dependent on the quality of the reference mirror. All results presented in this paper were obtained using a silicon wafer with a polished (111) plane as the reference mirror. The result of the observation was the creation of a fringe pattern, examples of which can be seen in Fig.~\ref{Fig1}(b) as well as in Fig.~\ref{Fig6}. 

\subsubsection{Phase-shifting method of phase calculations}

In two-beam interferometry, the electric fields of the reference wave and the wave from the sample summed up at the detector:

\begin{equation}
\bm{E}(x,y,t)= \bm{E}_1(x,y)e^{i(\omega t+\phi_1(x,y))}+\bm{E}_2(x,y)e^{i(\omega t+\phi_2 (x,y))}.
\end{equation}

The detector registers the intensity of the wave, which is proportional to the square of the total electric field:

\begin{equation}
I(x,y)\propto|\bm{E}(x,y,t)|^2=|\bm{E}_1(x,y)|^2+|\bm{E}_2(x,y)|^2+E_{12}(x,y)+E_{12}^*(x,y),   
\end{equation}

\noindent where  \(E_{12}(x,y)=\bm{E}_1(x,y)\bm{E}_2^*(x,y)e^{i\Phi(x,y)}\) and \(\Phi(x,y) = \phi_1 (x,y)-\phi_2 (x,y) \).

Neglecting the difference in the polarizations of the two beams, we obtain:

\begin{equation}
	I(x,y)=B(x,y)+A(x,y)\cos (\Phi(x,y)).	
\end{equation}

In most applications, all useful information is contained in the \(\Phi(x,y)\) phase.

To calculate \(\Phi(x,y)\) we used the method of phase-shifting interferometry. In this method different phase shifts $\delta$ are obtained by displacing the reference mirror. As a result of this shift we have:

\begin{eqnarray}
\nonumber I'(x,y)&=&B(x,y)+A(x,y)\cos (\Phi (x,y)+\delta)=B(x,y)+\\
~&~&A(x,y)(\cos \Phi (x,y)  \cos \delta-\sin \Phi (x,y) \sin \delta).
\end{eqnarray}

From a series of such measurements, one can obtain the desired phase. For example, from a series of six measurements with phase shifts $\delta_1=0$, $\delta_2=\pi/2$, $\delta_3=\pi$, $\delta_4=3\pi/2$, $\delta_5=2\pi$, $\delta_6=5\pi/2$, we get:

\begin{eqnarray}
I_1 (x,y)&=&B(x,y)+A(x,y)\cos (\Phi (x,y)),\\ 	 		
I_2 (x,y)&=&B(x,y)-A(x,y)\sin (\Phi (x,y)),\\  	 		
I_3 (x,y)&=&B(x,y)-A(x,y)\cos (\Phi (x,y)),\\ 	 		
I_4 (x,y)&=&B(x,y)+A(x,y)\sin (\Phi (x,y)),\\
I_5 (x,y)&=&B(x,y)+A(x,y)\cos (\Phi (x,y)),\\ 	 		
I_6 (x,y)&=&B(x,y)-A(x,y)\sin (\Phi (x,y)),  	 		
\end{eqnarray}

From the above formulas it is clear that $I_5 (x,y)=I_1 (x,y)$ and $I_6 (x,y)=I_2 (x,y)$. This is not entirely true because $A(x,y)$ and $B(x,y)$ may change slightly when the reference mirror is moved. However, by achieving maximum agreement between $I_1$ and $I_5$, as well as $I_2$ and $I_6$, we can obtain the best calibration. 

We can calculate tangent of $\Phi (x,y)$ from above equations using several different formulas, for instance it is easy to check that:

\begin{equation}
\tan [\Phi(x,y)]=\frac{I_4-I_2}{I_1-I_3}.\label{f4A}
\end{equation}

However, it turns out that different phase calculation formulas have different sensitivity to linear errors arising, for example, from calibration errors. In our work, we used the formula from the article by J.~Schmit and K.~Creath \cite{Schmit1995}, which is the most stable with respect to linear errors:
\begin{equation}
\tan [\Phi(x,y)]=\frac{I_1-5I_2-2I_3+10I_4-3I_5-I_6}{I_1+3I_2-10I_3+2I_4+5I_5-I_6}.\label{f6B}
\end{equation}

\begin{figure}[ht!]
\centering\includegraphics[width=0.95\textwidth]{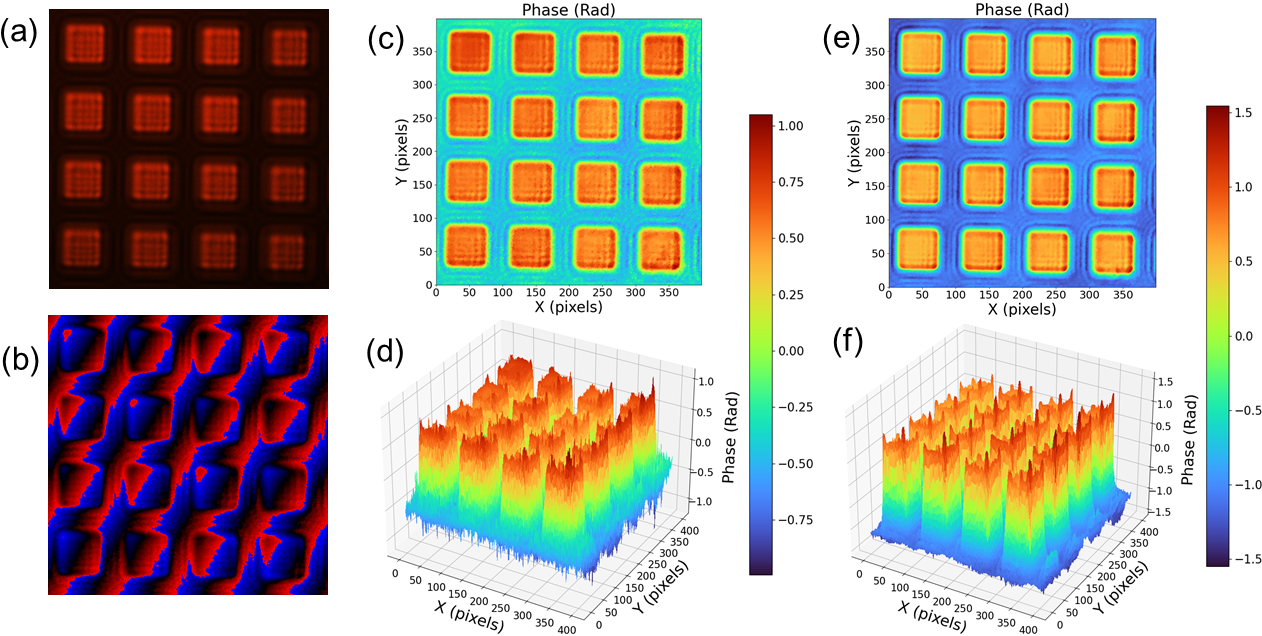} \caption{{Observation of an AFM (Atom Force Microscope) Height Standard, 100~nm. (a) View under a microscope in reflective light; (b)  phase $\Phi(x,y)$, calculated using the formula (\ref{f6B}); (c) and (e) are the contour graphics (d) and (f) are the 3D images. The X and Y axes show pixel numbers. The phase magnitude in (b) is depicted in false color. Positive phase is shown in red and negative phase is shown in blue. The brightness of the color is proportional to the absolute value of the phase. The Z axis in (d) and (f) and the color in (c) --  (f) show the phase $\Phi(x,y)$ in radians. The color scale is the same for (c), (d), as well as for (e) and (f), as indicated in the scales on the right. The data presented in Figures (a) through (d) were obtained using an LED with a center wavelength of 595~nm and a 600-nm filter with a 10-nm bandwidth. The data presented in Figures (e) and (f) were obtained using an LED with a center wavelength of 470~nm and a 470-nm filter with a 10-nm bandwidth.
\label{Fig2}}} 
\end{figure}

\subsubsection{Testing}

In Fig.~\ref{Fig2} we show the results of observation of 100~nm AFM Height Standard.  According to the manufacturer (TED PELLA, INC), the sample is a set of sqoare columns 100~nm high, located at intervals of 10~$\mu$m. From the set of six fringe patterns we calculated of $\Phi(x,y)$ phase with help of the equation (\ref{f6B}). The results are shown in Fig.~\ref{Fig2}(b). The problem is that since the tangent is a discontinuous function, the two-dimensional function $\Phi(x,y)$ we obtained is also discontinuous. In order to obtain the profile of the reflective surface of the sample, we need to find areas of continuity of the function $\Phi(x,y)$, number them in the required sequence, then for each point on the surface, shift the function $\Phi(x,y)$ by the required number of $\pi$ and, finally, level the resulting surface. Often this procedure is very difficult and sometimes impossible, but not in this case. The surface of the sample obtained in this way is shown in the figures (c) through (f).

The lateral calibration can be readily performed in accordance with Fig.~\ref{Fig2}. With regard to the X and Y scale, 10~$\mu$m is approximately equivalent to 100 pixels. Consequently, the scale unit in these circumstances is approximately 0.1~$\mu$m per pixel. In the Z-direction, the phase shift, $\Delta\Phi$, can be estimated when the height is changed by $\Delta h$ as follows: $\Delta\Phi = (2\pi\times 2\Delta h)/\lambda$, where the factor of two in $2\Delta h$ is due to the fact that in Linnik's interferometer, light travels this distance $\Delta h$ twice. At a height difference of 100~nm, we obtain a phase shift of 2.09~Rad for a wavelength of 600~nm and 2.67~Rad for a wavelength of 470~nm. As illustrated in Fig.~\ref{Fig2}, the measured surface of the sample is a rather intricate function. However, if we measure from the flat top of the column to its base, we obtain approximately 2.6~Rad for a light source with a wavelength of 470~nm and $\approx 1.7$~Rad for a light source with a wavelength of 600~nm. The discrepancy between the measured values and the calculated values is likely due to diffraction effects, as is the complex shape of the surface. This discrepancy decreases as the wavelength of the light source decreases. In the following section, we will utilize calculated values. However, it is essential to recognize that diffraction effects can introduce distortions, which must be accounted for in quantitative measurements on a case-by-case basis. After considering all the information provided, we can estimate the vertical spatial resolution of our method to be approximately 10~nm.

\subsection{The single-shot method with the polarize-sensitive camera}

\begin{figure}[ht!]
\centering\includegraphics[width=0.95\textwidth]{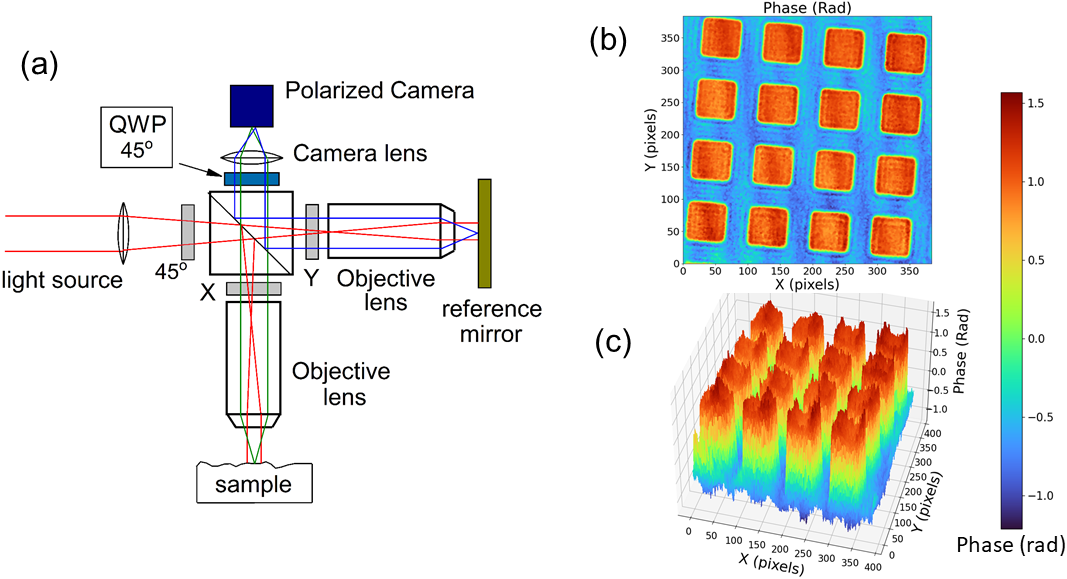} \caption{{Microscope scheme for single-shot detection and  results for  100 nm AFM Height Standard. (a) Modified Linnik interferometer. The light emitted by the LED passes through a linear polarizer oriented at an angle of 45 degrees. The symbols X and Y represent linear polarizers with axes oriented at 0 and 90 degrees, respectively. The quarter-wave plate (QWP) has its axes oriented at a 45-degree angle. (b)  Contour graphic. (c) 3D image.  The color scale is the same for (b) and (c). The image acquisition time was 3~ms.
\label{Fig3}}} 
\end{figure}

Fig.~\ref{Fig3}(a) shows the scheme of the modified Linnik interferometer for single shot interference pattern detection. We replaced the camera and added two linear polarizers and a quarter wave plate (QWP). The polarizers are perpendicular to each other so that the reflected light from the sample and the reference mirror do not interfere. The QWP is oriented at an angle of $45^\circ$ so that the light reflected from the sample and the reference mirror are circularly polarized in opposite directions. The polarization sensitive camera can simultaneously detect four images with different polarization $0^\circ$, $45^\circ$, $90^\circ$ and $135^\circ$. The two circularly polarized waves interfere with each other, and a phase shift equal to twice the camera shift occurs between the interference patterns for different images.  So from one shot we have four images with phase shifts $\delta_1=0$, $\delta_2=\pi/2$, $\delta_3=\pi$, $\delta_4=3\pi/2$. Using the formula (\ref{f4A}), we can easily obtain the tangent of the phase shift $\Phi(x,y)$ between the reference mirror and the sample under study. Next, we obtain the desired image of $\Phi(x,y)$ using the procedure described above. Similar but slightly more complex methods have been used in previous works \cite{Wyant2018,Millerd2001,Novak2005,Millerd2005,Brock2005,Millerd2006,Millerd2007,Hong2024} for Twyman-Green,  Fizeau, and Michelson interferometers.

The reliability of the above reasoning is easily confirmed using Jones calculus \cite{Fowles1989}. The total electric field of two waves reflected from the sample and the reference mirror with horizontal and vertical polarization, respectively, is equal to:

\begin{equation}
	E(x,y,t) =  E_1(x,y)\left({\begin{array}{cc}
	     1 \\
	     0 
	\end{array} }\right) \exp{[i(kz-\omega t +\phi_1)]}+
                   E_2(x,y)\left({\begin{array}{cc}
	     0 \\
	     1 
	\end{array} }\right) \exp{[i(kz-\omega t +\phi_2)]}. 
\end{equation}

After passing through the QWP oriented at a 45-degree angle, these two waves acquire opposite circular polarizations:

\begin{eqnarray}
\nonumber E'(x,y,t) =	\frac{1}{\sqrt{2}}\left({\begin{array}{cc}
	     1 & i \\
	     i & 1 
	\end{array} }\right) E(x,y,t) =
 \frac{E_1(x,y)}{\sqrt{2}}\left({\begin{array}{cc}
	     1 \\
	     i 
	\end{array} }\right) \exp{[i(kz-\omega t +\phi_1)]}+\\
 \frac{E_2(x,y)}{\sqrt{2}}\left({\begin{array}{cc}
	     1 \\
	     -i 
	\end{array} }\right) \exp{[i(kz-\omega t +\phi_2+\pi/2)]}.
\end{eqnarray}

Finally, after a linear polarizer with an axis of transmission angle of $\theta$ from the horizontal, we have:

\begin{eqnarray}
\nonumber E''(x,y,t) = \left({\begin{array}{cc}
	     \cos{^2(\theta)} & \cos{(\theta)}\sin{(\theta)} \\
	     \cos{(\theta)}\sin{(\theta)} & \sin{^2(\theta)}
	\end{array} }\right) E'(x,y,t) =\\
 \nonumber \frac{E_1(x,y)}{\sqrt{2}}\left({\begin{array}{cc}
	     \cos{(\theta)} \\
	     \sin{(\theta)} 
	\end{array} }\right) \exp{[i(kz-\omega t +\phi_1+\theta)]}+\\
 \frac{E_2(x,y)}{\sqrt{2}}\left({\begin{array}{cc}
	     \cos{(\theta)} \\
	     \sin{(\theta)} 
	\end{array} }\right) \exp{[i(kz-\omega t +\phi_2+\pi/2-\theta)]}
\end{eqnarray}

Then we can easily calculate the intensity of light:

\begin{equation}
I(x,y) \propto E''^{\dagger}(x,y,t) E''(x,y,t) =	\frac{1}{2}[|E_1(x,y)|^2 +|E_2(x,y)|^2 +E_{12}(x,y)+E^*_{12}(x,y)],
\end{equation}
where ${\dagger}$ means transpose and complex conjugation and 
\begin{equation}
E_{12}(x,y)=E_1(x,y)E^*_2(x,y)\exp{[i(\phi_1-\phi_2-\pi/2+2\theta)]}.
\end{equation}

As result we have:
\begin{equation}
I(x,y)=B(x,y)+A(x,y)\cos{ (\Phi(x,y)+2\theta)},
\end{equation}
where the unimportant constant shift $\pi/2$ is included in $\Phi(x,y)$.

We used a Phoenix IMX264MZR 5~MP Polarized Camera in our experiments capable of capturing imagees with polarizations of 0, 45, 90, and 135 degrees simultaneously.
The light source utilized was an LED with a central wavelength of 470~nm. The light passed through a 470~nm filter with 10~nm bandwidths and a polarizer oriented at an angle of 45 degrees. In the absence of a 45-degree input pulolarizer, no interference is observed. Parasitic reflections with a 45-degree polarization pass through the QWP unchanged, resulting in different image intensities at different polarizations. However, these unwanted effect can be significantly reduced by using an appropriate illumination tint to separate images of the sample and stray reflective surfaces.

Figures \ref{Fig3}(b) and \ref{Fig3}(c) display images of the 100 nm AFM Height Standard, identical to the one shown in figure \ref{Fig2}. The results were obtained by processing a single-shot image with an acquisition time of approximately 3~ms. As illustrated in the accompanying figures, the image quality is comparable to that obtained by the phase-shifting method. Additionally, experiments were conducted with laser light sources, yet the occurrence of parasitic interference, as a result of the long coherence length, led to a notable deterioration in image quality and, consequently, a reduction in the resolution of the microscope.

\section{Applications of interference microscopy method}

In this section, only the results obtained using the phase-shifting method with an LED light source with a center wavelength $\lambda_0 = 595$~nm and a line width at half maximum $\Delta \lambda \approx 70$~nm will be considered.

\subsection{Diamond surface}

\begin{figure}[ht!]
\centering\includegraphics[width=0.95\textwidth]{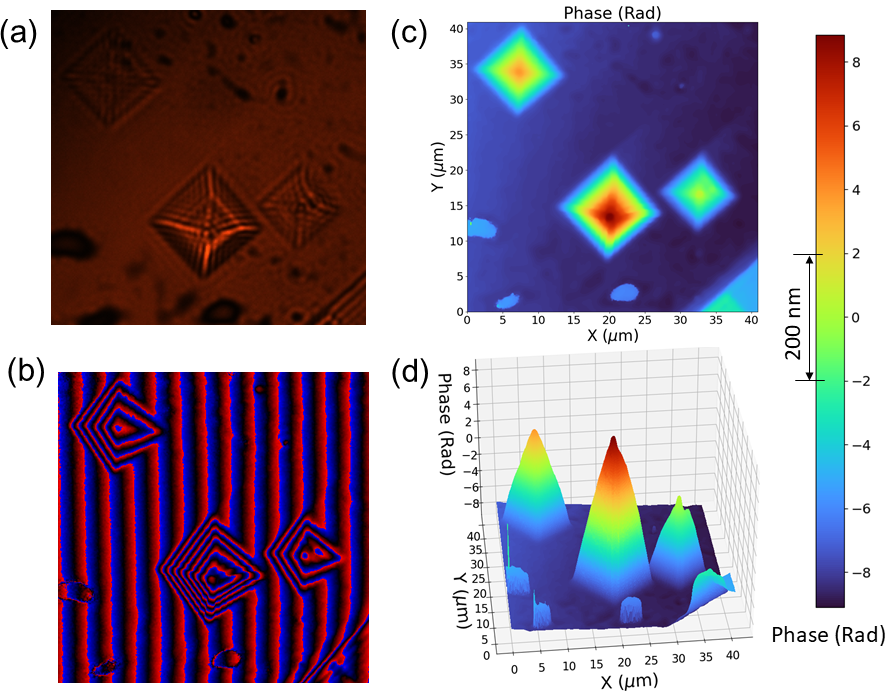} \caption{{Diamond surface. (a) View under a microscope in reflective light; (b) phase $\Phi(x,y)$, calculated using the formula (\ref{f6B}); (c) contour graphic; (d) 3D image. \label{Fig4}}} 
\end{figure}

Fig.~\ref{Fig4} shows the results of an experiment on observing the surface of an artificial diamond. Using a microscope, we can see protrusions on the surface of a diamond in the form of rectangular pyramids. Our method confidently determines the height and form of the protrusions.

Decoding the interferogram may become increasingly challenging as the angle of inclination of the slope of the 'pyramid' increases, resulting in narrower regions of phase continuity. Beyond a certain angle of inclination, the lateral resolution may become insufficient, which could make decoding more difficult.

\subsection{Biological objects}

\begin{figure}[ht!]
\centering\includegraphics[width=0.5\textwidth]{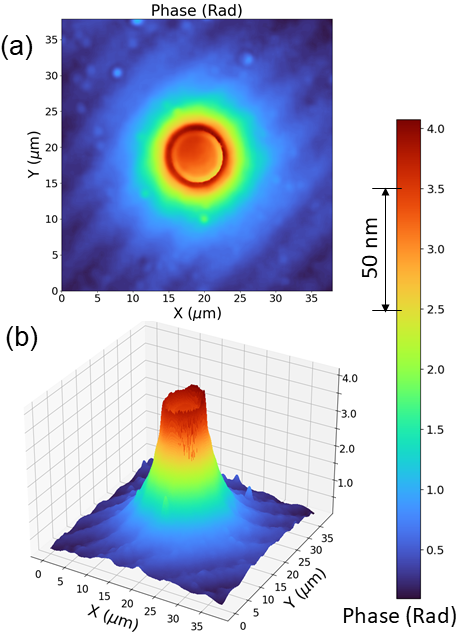} \caption{{Red blood cell (RBC).  (a) Contour graphic, (b) 3D image. \label{Fig5}}} 
\end{figure}

Fig.~\ref{Fig5} shows results of erythrocyte (RBC) observation. The blood was applied in a thin layer to a glass slide and observed immediately without the use of a coverslip. Thus, in this figure we observe the reflective surface of the red blood cell, slightly uppered above the surface of the liquid. It is necessary to take into account that the vertical scale is completely different than in the plane. The diameter of the red blood cell itself is about 7 $\mu$m, and the part protruding above the surface of the liquid is about 0.3 $\mu$m.

\section{Discussion}

\begin{figure}[ht!]
\centering\includegraphics[width=0.8\textwidth]{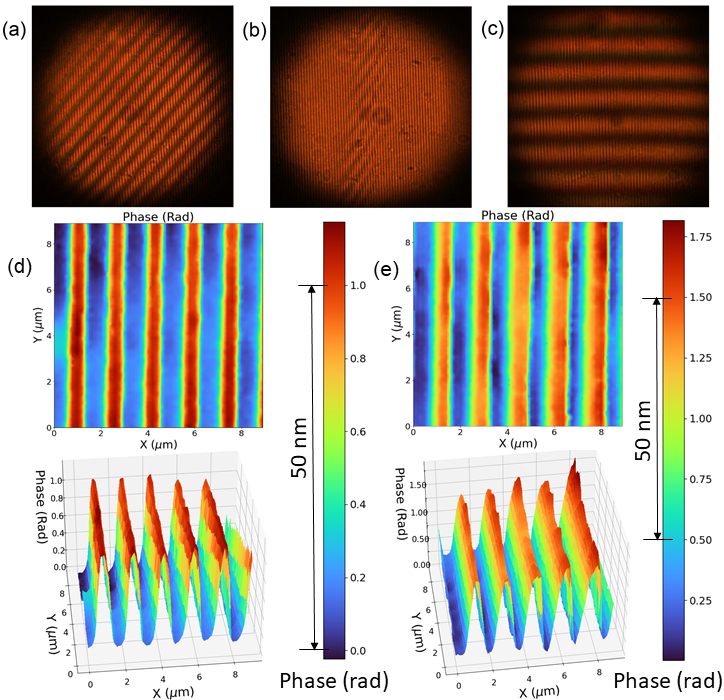} \caption{{Fringed pattern of holographic reflective grating 600/mm with different orientations of sample and reference mirror. (a) The grating and the mirror were located almost perpendicular to the optical axis; (b) the grating was rotated approximately 15 degrees; (c) the reference mirror was rotated to the same angle ($\approx 15~{\rm DEG}$) as the grating. (d) Contour graphics in the top row and 3D images in the bottom of the grating and the mirror were located almost perpendicular to the optical axis; (e) the same for both the grating and the mirror were rotated to the same angle approximately 15~DEG.
\label{Fig6}}} 
\end{figure}

Let's discuss the advantages that Linnik interferometer provides. The symmetrical design of the interferometer makes it easy to operate with incoherent light sources, as in our case. In addition, chromatic aberrations are eliminated. Also, the symmetrical design allows for quick replacement of one pair of equivalent objective lenses with another with minimal adjustment of the optical system.  The greatest contrast in the interference pattern is achieved when the intensity of light from the sample and the reference mirror is equal. In the Linnik interferometer, a simple procedure for replacing the reference mirror allows one to obtain optimal conditions for observing interference fringes.

By tilting the reference mirror, you can achieve the best possible appearance of the interference pattern for further phase calculations. It turns out that when using objective lenses with a large numerical aperture, tilting the reference mirror by a significant angle can compensate for the non-optimal position of the sample.

Fig.~\ref{Fig6} shows the interference pattern for various positions of the sample and the reference mirror. The sample here was a reflective grating. In Fig.~\ref{Fig6}(a) it was located perpendicular to the optical axis, but in the figures \ref{Fig6}(b) and \ref{Fig6}(c) it was tilted by 15 degrees. The reference mirror was located perpendicular to the optical axis in the figures \ref{Fig6}(a) and \ref{Fig6}(b). In the Fig~\ref{Fig6}(c) the mirror was turned to the exactly the same angle as the grating and the lighting was adjusted accordingly. 

The fringe pattern in Fig.~\ref{Fig6}(a) enables accurate surface shape calculation, as shown in Fig.~\ref{Fig6}(d). Rotating the sample reveals a narrow strip with a complex internal structure in the fringe pattern. The width of this strip is determined by the coherence length of the light. Deciphering the image in this instance is impossible. Rotating the reference mirror to the same angle as the sample produces a fringe pattern suitable for surface calculations (see Fig.~\ref{Fig6}(e)). The calculated grating profiles are nearly identical in both cases.

The ability to observe the shape of a surface at a significant angle to the lens axis with super-resolution provides extensive opportunities for studying objects with non-flat shapes, such as a convex hemisphere, without having to tilt them. When scanning such surfaces, it is only necessary to tilt the reference mirror to the desired angle and adjust the lighting system.

The potential for enhancing measurements by tilting the reference mirror for a profilometer based on a Michelson interferometer has been previously investigated \cite{Wu2024}. However, due to the small numerical aperture (NA) of the objective lens, the angle of inclination of the reference mirror was constrained to a minimal degree (no more than $0.39^\circ$). In contrast, due to the high NA, we can operate with considerably larger tilt angles, which presents a significant opportunity.

The phase-shifting method of phase calculation we used is a simple method that gives reliable results. However, it is not without its drawbacks. The need to obtain several images and a long experimental time makes it impossible to observe moving objects and causes a high sensitivity to vibrations. The small drift of the interference pattern is not so significant.  The optimal formula for calculating the phase (\ref{f6B}) effectively eliminates the linear errors it introduces. However, vibration causes significant distortion in the phase calculation results. We used two-stage pneumatic vibration isolation method was utilized to reduce vibration to an acceptable level.

The Linnik interferometer was adapted for single-shot measurements by using a polarized camera. This modification makes the microscope almost insensitive to vibrations and allows the observation of moving objects. However, the spatial resolution of a polarized camera is halved, which significantly degrades the resulting image quality and the longitudinal resolution of the microscope.  Nevertheless, we have identified potential solutions to these issues, which should significantly enhance the microscope's efficiency in single-shot mode.

\section{Conclusion}

We have designed and built a microscope based on the Linnik interferometer using two methods of interference pattern measurement. This configuration of the microscope makes it easy to change objective lenses and observe a variety of samples. With the phase-shifting method, we have a spatial resolution perpendicular to the sample surface of about 10~nm. In addition, this normal to the surface can deviate at a significant angle from the axis of the objective. The lateral resolution is the same diffraction limited as that of a conventional microscope, taking into account the limitation of the tilt angle of the surface structure details to obtain an appropriate interference fringe pattern. The single-shot method is insensitive to vibration and has great prospects for improvement.

\section*{Acknowledgments}

The work was supported by the Keck Foundation.

\end{document}